\documentclass[12pt,letterpaper,a4paper]{article}  

\usepackage[includeheadfoot,
            marginratio={1:1,2:3}, 
            width=412pt, 
            height=688pt,]{geometry}

\usepackage{amsmath}
\usepackage{amsfonts}
\usepackage{amssymb}

\usepackage{empheq}

\newcommand{\nc}{\newcommand}
\nc{\beq}{\begin{equation}}
\nc{\eeq}{\end{equation}}
\nc{\bea}{\begin{eqnarray}}
\nc{\eea}{\end{eqnarray}}

\newcommand{\eq}[1]{\begin{equation}
                     \begin{split} #1 \end{split}
                     \end{equation}}
                     
\newcommand{\bom}[1]{\fboxsep2.5mm\fbox{
           $ \displaystyle{ #1} $}}

\begin{document}

\vspace*{-1.5cm}
\begin{flushright}
  {\small
  MPP-2012-10\\
  ITP-UU-12/07\\
  SPIN-12/06\\ 
  }
\end{flushright}

\vspace{1.5cm}
\begin{center}
  {\LARGE
Palatini-Lovelock-Cartan Gravity -- \\[0.25cm]
Bianchi Identities for Stringy Fluxes
}

\end{center}

\vspace{0.75cm}
\begin{center}
  Ralph Blumenhagen$^{1}$, Andreas Deser$^{1}$, Erik Plauschinn$^{2}$ and 
Felix Rennecke$^{1}$
\end{center}

\vspace{0.1cm}
\begin{center} 
\emph{$^{1}$ Max-Planck-Institut f\"ur Physik (Werner-Heisenberg-Institut), \\ 
   F\"ohringer Ring 6,  80805 M\"unchen, Germany } \\[0.1cm] 
\vspace{0.25cm}
\emph{$^{2}$ Institute for Theoretical Physics and Spinoza Institute, \\
Utrecht University, 3508 TD  Utrecht, The Netherlands}  \\

\vspace{0.2cm}

 \vspace{0.5cm} 
\end{center} 

\vspace{1cm}


\begin{abstract}
A Palatini-type action for Einstein and Gauss-Bonnet gravity
with non-trivial  torsion is proposed. 
Three-form flux is incorporated via a deformation of the 
Riemann tensor, and consistency of the Palatini variational principle 
requires  the flux to be covariantly constant and to satisfy a Jacobi identity.
Studying gravity actions of third order in the curvature leads 
to a  conjecture about general 
Palatini-Lovelock-Cartan gravity. We point out 
potential relations to  string-theoretic Bianchi identities
and, using the Schouten-Nijenhuis bracket, 
derive a set of Bianchi identities for the
non-geometric $Q$- and $R$-fluxes which include derivative
and curvature terms.
Finally, the problem of relating torsional gravity 
to higher-order corrections
of the bosonic string-effective action is revisited.
\end{abstract}

\clearpage



\section{Introduction}
\label{sec:intro}

One of the most distinctive and generic features of 
string theory and supergravity   
is that at the massless level gravity is extended by 
an axionic anti-symmetric two-form field and the dilaton.
In the case of string theory, this is a direct consequence of quantizing the 
string excitations.
The Kalb-Ramond field and its contribution to the
low-energy effective action have  been a subject of study throughout
the history of string theory. 
In particular, the application of  T-duality 
\cite{Shelton:2005cf,Wecht:2007wu}
led to a picture where the three-form flux of the Kalb-Ramond field
was dualized not only to geometric flux, but also to so-called
non-geometric fluxes. Here one distinguishes the still locally geometric
$Q$-flux\footnote{See for instance  the analysis of \cite{Andriot:2011uh}
for an effective action at leading order for the case of $Q$-flux.}
from the so-called $R$-flux, which is not even locally geometric. The former
gave rise to the notion of $T$-folds whereas the  
latter was argued to be related to a nonassociative  geometry
\cite{Bouwknegt:2004ap,Blumenhagen:2010hj,Lust:2010iy,Blumenhagen:2011ph}.
Furthermore, assuming that T-duality is not a symmetry only of certain solutions
of string theory but is a symmetry of the theory itself led
to the idea of double field theory \cite{Hull:2009mi,Hull:2009zb}. 
In this approach, one doubles the 
coordinates and formulates an effective action which
is invariant under the T-duality group $O(D,D)$
in $D$-dimensions.
This formalism was very successful for the 
action at leading order in a derivative expansion, 
but was shown to become more involved
at next to leading order \cite{Hohm:2011si}.

The purpose of this paper is to approach the question 
about  the nature of the Kalb-Ramond
field and its T-dual incarnations from a more  gravity
based direction.
Long ago it was realized \cite{Scherk:1974mc} that at leading order 
in the string tension
$\alpha'$, i.e. at the two-derivative level, the low-energy effective action
for  the graviton and the Kalb-Ramond field $B$ 
 follows from  Einstein  gravity with
a connection whose  torsion is  equal to $H=dB$ -- also known as Einstein-Cartan theory. 
However, already at second order in $\alpha'$ this geometric
picture was shown to break down. In particular, for vanishing $H$-flux the on-shell
string scattering amplitudes of the graviton are known to be consistent with 
the  ghost-free Gauss-Bonnet action. But for non-vanishing flux,
in \cite{Bern:1987wz} it was shown 
 that the latter action with the torsional connection
is not consistent with the string equations of motion.

In the work mentioned above  the choice of  connection was
put in by hand and did not follow dynamically from an equation
of motion. For the Einstein gravity case there exists an action 
which remedies this  point, the  so-called 
Einstein-Palatini action, whose Lagrangian density  is considered 
to be a functional  both of the metric and the connection \cite{17604}
(see also, for instance, \cite{Arnowitt:1962hi}).
The field equations following from the variation with respect to the connection 
then imply that the latter has to be Levi-Civita. 
In \cite{Exirifard:2007da} 
(see also  \cite{Borunda:2008kf,BasteroGil:2009cn}) 
it was shown that this relation extends 
to the Gauss-Bonnet and, in fact, to all higher-order Lovelock gravity 
actions  \cite{Lovelock:1971yv}.
(See also \cite{Lanczos:1938sf} for earlier work on higher-order gravity actions.)
This remarkable result  relies on the fact that
 for the Lovelock-type  combination of higher order
curvature terms  the Bianchi identities for the curvature tensor
lead to non-trivial cancellations. 
In \cite{Dadhich:2010dg} it was pointed out that the Lovelock actions are also
singled out by the fact that they lead to consistent truncations. 
This means that the equations of motion in the Palatini approach are equivalent to
those resulting from the variation of the Lovelock action
with the Levi-Civita connection inserted by hand.

In view of this situation, we  ask the question whether
the Palatini formalism can be generalized to include the case where
the torsion does not vanish and is identified with
the field strength of the Kalb-Ramond field.
More concretely, we seek for a Palatini-type torsional 
Einstein-Hilbert, Gauss-Bonnet or even general Lovelock action -- a Palatini-Lovelock-Cartan action --
which on-shell reduces to the
corresponding  action with torsion\footnote{For an approach to Lovelock-Cartan theory see \cite{Mardones:1990qc}.}.
Deforming the curvature tensor by Lagrangian multipliers of the form 
\ {\it three-form}$\times${\it torsion}\,, we 
 find that this is indeed possible. And, as mentioned above, since the
curvature Bianchi identities were playing the key role in the
non-torsional case, one could expect that the Bianchi identity
for the Kalb-Ramond field  will be equally important. 
This is true but, as we will deduce
in detail, the latter Bianchi identity turns out  not to be  sufficient.
Instead, our computation points towards   the   stronger conditions that
the three-form is covariantly constant and satisfies
a Jacobi identity.

We observe that these conditions are not unfamiliar from string
theory. They can be considered as
the basic two requirements
guaranteeing that also all  Bianchi identities in the geometric
as well as non-geometric T-dual descriptions are satisfied.
As a new result, we derive  Bianchi identities for the
non-geometric $Q$- and $R$-flux not just for constant
fluxes but with curvature terms and 
covariant derivatives of the fluxes included.
For this purpose, we exploit the  definition of the $R$-flux
in terms of the Schouten-Nijenhuis bracket of a bi-vectorfield.

Moreover, we revisit the question whether the Gauss-Bonnet action
with torsion does  reproduce the on-shell string scattering amplitudes
for  the graviton and the Kalb-Ramond field (for vanishing dilaton).
Recall that in \cite{Bern:1987wz} a negative  
answer was given. There, of course,
only the Bianchi identity $dH=0$ was taken into account to
relate the various possible diffeomorphism invariant combinations
in the action. We show that if  one uses in addition the two
stronger conditions we derived from the Palatini approach,
then the pure Gauss-Bonnet action with torsion remains to be  in conflict with
the string constraints. Nevertheless, we show that
one can write the string corrections in  a Gauss-Bonnet
type form.

This paper is organized as follows: in section \ref{sec_linear}, we 
review some aspects of the Einstein-Palatini action and
generalize it to include torsion.  In section \ref{sec:GB}, we analyze
the Palatini approach to the Gauss-Bonnet action with torsion and
obtain two new restrictions.
In section \ref{sec:lovelock}, we revisit the Gauss-Bonnet action in view of 
the new conditions, and show  that this formalism is consistent
also for the third-order Lovelock action. The section closes 
with our central  result, namely 
a  conjecture about Palatini-Lovelock-Cartan gravity. 
In section \ref{sec:ST} we connect our findings to the string theory action, and
section \ref{sec:bianchi}
contains  a  detailed analysis of the Bianchi identities 
for three-form fluxes in string theory.
Readers mainly interested in these results, may directly move
to this section.
In section \ref{sec:con} we conclude
and discuss some  open questions.


\section{Einstein-Palatini gravity with torsion}
\label{sec_linear}

In this section, we deform Einstein gravity 
by a completely anti-symmetric three-index object
$\eta_{abc}(x)=\partial_a \beta_{bc}(x)+\partial_b \beta_{ca}(x)+\partial_c \beta_{ab}(x)$. Our aim is to obtain a theory
with torsion, whose field equations are isomorphic
to the leading order string equations of motion
for the metric and Kalb-Ramond field in the case of a 
constant dilaton.

Our approach is to work with the Einstein-Palatini formalism in which the metric $g$ as well as the connection $\Gamma$
are treated as independent fields. By solving the field equation for $\Gamma$, one is led to 
a metric-compatible and torsion-free connection and thus recovers the Einstein-Hilbert theory.
In the following, we recall this formalism for usual Einstein gravity and introduce our
basic conventions.


\subsubsection*{Einstein-Palatini action}

The Einstein-Palatini  action in $n$ space-time dimensions is given by 
\eq{
\label{actionEP}
  \mathcal S= \frac{1}{2\kappa^2} \int d^nx\; \sqrt{-g}\, g^{ab}\,
  R_{ab}(\Gamma) \;,
}
where the Lagrangian density ${\cal L}(g^{ab},\Gamma^a{}_{bc})$ 
is a functional of the inverse metric $g^{ab}$ and the connection 
$\Gamma^a{}_{bc}$. 
The Riemann curvature tensor, and therefore also the Ricci tensor $R_{ab}(\Gamma)$, only depends on the connection $\Gamma$ 
\eq{
    R^a{}_{bcd}=\partial_c \Gamma^a{}_{db}-\partial_d
    \Gamma^a{}_{cb}+\Gamma^a{}_{c m}\,
    \Gamma^m{}_{db}-\Gamma^a{}_{d m}\, \Gamma^m{}_{cb}  \;.
}
The variation of the Riemann curvature tensor with respect to the connection is given by the Palatini formula 
\eq{
  \label{var_palatini}
    \delta R^a{}_{bcd} =\nabla_c (\delta
    \Gamma^a{}_{db})-\nabla_d (\delta
    \Gamma^a{}_{cb}) + T^m{}_{cd} (\delta
    \Gamma^a{}_{mb})\; ,
}       
where $T^a{}_{bc}=\Gamma^a{}_{bc}-\Gamma^a{}_{cb}$ denotes the torsion tensor.
The covariant derivative $\nabla_{c}$ appearing in \eqref{var_palatini}
acts on $(1,1)$-tensors as follows
\eq{
       \nabla_{c}  t^{a}{}_{b}=
     \partial_c t^{a}{}_{b}
    -\Gamma^m{}_{cb}\; t^{a}{}_{m} + \Gamma^a{}_{c m}\; t^{m}{}_{b}
    \;,
}
with a straightforward generalization to  $(p,q)$-tensors.

Furthermore, as will become clear in the following, we are interested in metric-compatible connections which satisfy $\nabla_{a} g_{bc}=0$. These can be expressed in terms of the
Levi-Civita connection $\mathring \Gamma^a{}_{bc}$ and the contorsion $K^a{}_{bc}$ as follows
\eq{ 
  \Gamma^a{}_{bc}=  \mathring   \Gamma^a{}_{bc}  +     K^a{}_{bc} \;,
}
where the former, when expressed in terms of the metric, is given by
\eq{
  \label{christoph}
   \mathring \Gamma^a{}_{bc} =
    \frac{1}{ 2}\, g^{am}\left( \partial_b g_{m c} + \partial_c g_{bm } -
    \partial_m g_{b c} \right) \;.
} 
The contorsion can be written using the torsion tensor as 
\eq{
  \label{def_contorsion}
  K^{a}{}_{bc} = \frac{1}{2} \left( T^a{}_{bc} + T_b{}^a{}_c + T_c{}^a{}_b \right) \;.
}
Finally, we note that in case of a metric-compatible connection we can apply the following formula
for some vector $A^{a}$
\eq{
  \label{part_int_01}
    \sqrt{-g}\, \nabla_{m} A^{m} =\partial_{m}(\sqrt{-g}\, A^{m}) + \sqrt{-g}\,
     T^n{}_{nm}\, A^{m} \;.
}

Let us point out that in order to keep the length of the expressions to a bearable level,
throughout this paper we present formulas only up to terms which
vanish for metric-compatible connections.
Furthermore, we assume that the contorsion $K^a{}_{bc}$ is totally anti-symmetric, 
which also implies that the trace part of the torsion  $T^a{}_{ab}$ vanishes.
(In \cite{Dadhich:2010xa} it has been argued that the latter condition is equivalent to a gauge fixing.)
Choosing then for instance $A^a=B^{a
  m_1\ldots m_n}\, C_{m_1\ldots m_n}$ in equation \eqref{part_int_01},
we can perform an integration by parts to obtain the relation
\eq{
\int d^nx\, \sqrt{-g}\, B^{a m_1\ldots m_n}\, \nabla_a
C_{m_1\ldots m_n}
=-\int d^nx\, \sqrt{-g}\, \nabla_a B^{a m_1\ldots m_n}\, 
C_{m_1\ldots m_n} \;.
}

The above formulas now allow us to easily compute the variation of the Einstein-Palatini action \eqref{actionEP} with respect to the
connection $\Gamma^\mu{}_{\nu\lambda}$. Up to terms proportional to $\nabla_a g_{bc}$ 
and $T^a{}_{ab}$ which vanish with our assumptions, we obtain
\eq{
  \label{connectionvary}
       T^{\lambda\nu}{}_\mu=0 \;.
}
We can therefore conclude that a metric-compatible connection with vanishing torsion 
is a solution to the equations of motions. 
This connection is uniquely given by the Levi-Civita connection \eqref{christoph}.
Inserting  this result into the Einstein-Palatini action \eqref{actionEP},  we arrive at the 
usual Einstein-Hilbert form of general relativity.
The resulting equations of motion for the metric are the same
as the ones arising from 
first varying  the Einstein-Palatini action \eqref{actionEP} and then inserting
the Levi-Civita connection. This non-trivial feature is called
a consistent truncation (see e.g. \cite{Dadhich:2010dg})
and can be visualized by the
following commuting diagram:
\eq{
   \begin{array}{ccc}    {\cal L}(g,\Gamma) & \xrightarrow{\;\mbox{\scriptsize trunc.}\;} 
   & \overline{\cal   L}(g)\\[-2mm]
     \rotatebox{-90}{$\longrightarrow$} & &      \rotatebox{-90}{$\longrightarrow$} \\[7mm]
    {\rm EOM}       & \xrightarrow{\;\mbox{\scriptsize trunc.}\;}  &  \overline{\rm EOM} 
\end{array}
}


\subsubsection*{Einstein-Palatini with torsion} 

Motivated by string theory, we now consider a theory which in addition to the
metric contains a Kalb-Ramond field with field strength $\eta_{abc}$.
Our goal is to find a deformation of the Einstein-Palatini
action so that the torsionful connection 
\eq{
\label{torsionalconn}
\Gamma^a{}_{bc}=\mathring \Gamma^a{}_{bc}+
C\, \eta^a{}_{bc}
}
is a solution to the equations of motion for $\Gamma$, and that 
 on-shell the action
reduces to the Einstein-Hilbert case for the torsional 
connection \eqref{torsionalconn}.
A guess for an action which satisfies these conditions is the following
\eq{
\label{actionEPtors}
  \mathcal  S= \frac{1}{2\kappa^2} \int d^nx\; \sqrt{-g}\, g^{ab}
  \left( R_{ab}(\Gamma) + C\, \eta_{b m}{}^n\, T^m{}_{n a}
   -2 C^2\, \eta_{a}{}^{mn}\, \eta_{b mn} \right)\; . 
}
Note that, since the three-tensor $\eta_{abc}$ is coupled to the
torsion tensor, the deformation  is  diffeomorphism invariant.

Performing the variation of  \eqref{actionEPtors} with respect to
the connection, 
we obtain an additional contribution to \eqref{connectionvary} due to the deformation.
In particular, with our assumptions above, the field equations for $\Gamma$ read
\eq{
  \label{exp_02_04}
 T^{\lambda\nu}{}_\mu-2C \eta^{\lambda\nu}{}_\mu=0 \;.
 }
Recalling \eqref{def_contorsion}, we see that this is indeed solved by \eqref{torsionalconn}. 
Moreover, the last term in the action
\eqref{actionEPtors} ensures that,
after employing \eqref{exp_02_04}, we are left with 
the Einstein-Hilbert action with the torsional connection 
\eqref{torsionalconn}.

The Riemann curvature tensor for the connection \eqref{torsionalconn} can expressed in quantities involving the Levi-Civita connection and $\eta_{abc}$ as follows
\eq{
\label{curvexpand}
    R^{a}{}_{bcd}(\Gamma)=\mathring R^{a}{}_{bcd} (\mathring \Gamma)
    +C\left( \mathring\nabla_c\,  \eta_{db}{}^{a} -\mathring\nabla_d\, 
  \eta_{cb}{}^{a}\right) - C^2 \left(\eta^{a}{}_{c}{}^m 
  \eta_{bd m}-\eta^{a}{}_{d}{}^m \eta_{bc m}\right) \;,
}
where $\mathring R^{a}{}_{bcd} (\mathring \Gamma)$ 
and $\mathring\nabla$ denote the curvature tensor 
and 
covariant derivative for the Levi-Civita connection $\mathring \Gamma$, respectively.
For the Ricci tensor and scalar we find
\eq{ 
  R_{ab}(\Gamma)&= \mathring R_{ab}(\mathring \Gamma) 
  -C\,  \mathring\nabla_m \eta_{ab}{}^m 
     -C^2\, \eta_a{}^{mn}\, \eta_{b mn} \;, \\[0.1cm]
     R(\Gamma)\hspace{9.5pt}&=\mathring R(\mathring \Gamma) -C^2\, \eta^{mnp}\, \eta_{mnp}\;,
}
so that  the truncated action \eqref{actionEPtors}
becomes
\eq{
\label{truncact}
   \mathcal S=\frac{1}{2\kappa^2} \int d^nx\; \sqrt{-g}\,\left(\mathring R
   -C^2\, \eta_{abc}\, \eta^{abc}  \right) \;.
}
Note that, by varying  the action with respect to the inverse
metric and Kalb-Ramond field $\beta$ (where $\eta=d\beta$), one can show
that \eqref{truncact} is a consistent truncation
of the Palatini-Einstein-Cartan action \eqref{actionEPtors}.
Furthermore, we observe that for $C^2=1/12$, expression \eqref{truncact} is identical to the leading order       
string-theoretic  gravity action for constant dilaton 
\eq{
  \label{stringaction}
  \mathcal S_{\rm string}= \frac{1}{2\kappa^2} \int d^nx\; \sqrt{-g}\,\left( \mathring R
   - \frac{1}{12}\, H_{abc}\, H^{abc}  \right) \, .
}

Of course, it is well-known that at leading order in $\alpha'$ the three-form flux can be 
interpreted as torsion of the underlying Riemannian geometry
\cite{Scherk:1974mc}. An interesting question is whether
this is also true at higher orders in $\alpha'$, which has been
analyzed at next to leading
order in \cite{Bern:1987wz}. However, as far as we know, the analysis has never been performed
in conjunction with the Palatini formalism. In the next section, we  study 
the Palatini formalism at  higher order.


\section{Palatini action for Gauss-Bonnet gravity}
\label{sec:GB}

The aim of this section is to mimic the setup and analysis of the previous  for gravity theories quadratic in the curvature. We consider Gauss-Bonnet gravity for its compatibility with the Palatini approach, which also provides a consistent truncation in the torsion-free case. These properties stem from those of Lovelock gravity which will be discussed in section \ref{sec:lovelock}.


\subsubsection*{The torsion-free case}

The Gauss-Bonnet action can be written as
\eq{  
  \label{action_gb}
  \mathcal S_{\rm GB}
  =\frac{\alpha'}{2\kappa^2}\int d^n x \sqrt{-g}\, \Bigl( {R}_{abcd}\, {R}^{cdab} - 
           { R}_{ab}\, {R}^{ba} + 2  
          {R}_{ab}\, \widetilde{R}^{ba} - 
          \widetilde{R}_{ab}\, \widetilde{R}^{ba} + {R}^2 \Bigr),
}
where $R_{ab}=R^m{}_{amb}$ and
$\widetilde R_{ab}=g^{mn}\,R_{a m n b}$. 
Note that for a
generic connection, the Riemann curvature tensor is anti-symmetric only in the last two
indices, and that the two different types of contractions for the Ricci tensor give different contributions to the equations of motion. However, for a metric-compatible connection also the first two indices of the Riemann tensor will be 
anti-symmetric.

Let us review the  Palatini formalism of the Gauss-Bonnet action, for which we will
employ the following notation. The variation of  \eqref{action_gb} with respect to $\Gamma^\mu{}_{\nu\lambda}$ will be written as
\eq{
 \label{def_variation} 
  \delta_{\Gamma} \mathcal S_{\rm GB} = \frac{\alpha'}{2\kappa^2} \int d^n x\, \sqrt{-g}
  \; 2\, \delta\Gamma^\mu{}_{\nu'\lambda'}\, g^{\nu\nu'}\, g^{\lambda\lambda'} F_{\mu\nu\lambda} \;,
}  
and the contribution of the five terms in  \eqref{action_gb}  to  $F_{\mu\nu\lambda}$, up to terms vanishing for $\nabla_a g_{bc}=0$ and $T^a{}_{ab}=0$, are listed below
\eq{
  \label{variatgaussbonnet}
  \renewcommand{\arraystretch}{1.75}
  \begin{array}{lcl}
   +{R}_{abcd}\, {R}^{cdab} &:& \displaystyle \Bigl[\nabla^m( {R}_{\nu m\mu\lambda}-{R}_{m\nu\mu\lambda}) + T_\nu{}^{mn}\,
   {R}_{mn\mu\lambda} \Bigr]  \;, \\
   -{R}_{ab}\, {R}^{ba} &:& \displaystyle \Bigl[ \nabla_\mu {R}_{\nu\lambda} - g_{\mu\nu}\,
    \nabla^m {R}_{m\lambda} - T_{\nu\mu}{}^m\, {R}_{m\lambda} \Bigr] \;, \\
   -\widetilde{R}_{ab}\, \widetilde{R}^{ba} &:& \displaystyle\Bigl[\nabla_\lambda \widetilde{R}_{\nu\mu} - g_{\nu\lambda}\,
    \nabla^m \widetilde{R}_{m\mu} - T_{\nu\lambda}{}^m\, \widetilde{R}_{m\mu} \Bigr] \;, \\
   +2{R}_{ab}\, \widetilde{R}^{ba} &:& \displaystyle \Bigl[-\nabla_\mu \widetilde{R}_{\nu\lambda} + g_{\mu\nu}\,
    \nabla^m \widetilde{R}_{m\lambda} + T_{\nu\mu}{}^m\, \widetilde{R}_{m\lambda} \\
   && \displaystyle \hspace{5pt}-\nabla_\lambda {R}_{\nu\mu} + g_{\nu\lambda}\,
    \nabla^m {R}_{m\mu} + T_{\nu\lambda}{}^m\, {R}_{m\mu}\Bigr]  \;, \\
  +{R}^2 &:& \displaystyle \Bigl[ g_{\mu\nu}\, \nabla_\lambda {R}- g_{\nu\lambda}\, \nabla_\mu {R} +T_{\nu\mu\lambda}\, {R} \Bigr]   
  \;.
   \end{array}
}
The equation of motion for the connection is the sum
of these five contributions being equal to zero. 
To show that this field equation is satisfied  for the Levi-Civita connection, 
we first realize that in this case  $ \widetilde{R}_{ab}= -{R}_{ba}$.
Next, we recall  the second Bianchi identity for the curvature $R_{abcd}$
(with vanishing torsion) which reads
\eq{
     \mathring\nabla_{[\underline{a}} \mathring R^{mn}{}_{\underline{bc}]}=0 \;.
}
Note that we underline the indices which are anti-symmetrized. We can then utilize
\eq{
   \label{Bianchiapply}
   \mathring \nabla^m \mathring R_{\nu m\mu\lambda}=-\mathring\nabla_\mu \mathring R_{\nu\lambda} + 
   \mathring\nabla_\lambda \mathring R_{\nu\mu}\;, 
   \hspace{60pt}
   \mathring \nabla^m \mathring R_{m\lambda}= \frac{1}{2} \mathring\nabla_\lambda \mathring R\; ,
}
to show that all terms in \eqref{variatgaussbonnet} which involve a covariant derivative
cancel. The torsion terms vanish since we consider the Levi-Civita connection.
Therefore, after performing the variation of the Gauss-Bonnet action
with respect to the connection, the resulting equation of motion
is satisfied for the Levi-Civita connection.

We also remark that  the next to leading
order correction for the bosonic-string action 
for vanishing three-form flux and dilaton can be cast
into the Gauss-Bonnet form. In fact, this form of the action is singled out
as it  is explicitly ghost free \cite{Zwiebach:1985uq}.


\subsubsection*{The torsional  case}

Our objection is to analyze whether the above result can be generalized
to the torsion-full connection \eqref{torsionalconn}. 
Since terms of the form $\nabla_{\star} R_{\star\star\star\star}$ again have to be cancelled,
similarly to the analysis in section \ref{sec_linear},
we expect a deformation of the curvature tensors and scalars to involve the three-index object
 $\eta_{abc}$.

However, in the present situation the computation will become more involved.
For instance, in the case of non-vanishing torsion the 
first and second Bianchi identity read
\begin{align}
     \label{bianchi_tor_01}
    &R^a{}_{[\underline{bcd}]} =   \nabla_{[\underline b}\, T^a{}_{\underline{cd}]}+
             T^a{}_{m [\underline{b}}\, T^m{}_{\underline{cd}]} \;,
             \\[0.2cm]
     \label{bianchi_tor_02}
     &\nabla_{[\underline{a}} R^{mn}{}_{\underline{bc}]}=
     T^p{}_{[\underline{ab}}\; R^{mn}{}_{\underline{c}]p} \;.
\end{align}
Therefore, anticipating the final result, we again make the assumptions 
that the full connection is metric compatible, 
implying for instance that $R_{abcd}=-R_{bacd}$, as well 
as that the torsion tensor is completely anti-symmetric. Of course, the whole analysis 
can be performed without these assumptions, but then the 
underlying structure is less obvious.

Let us now consider the first term in \eqref{variatgaussbonnet}. 
We employ \eqref{bianchi_tor_02} to write
\eq{
\label{appleone}
   \nabla^m R_{\nu m\mu\lambda}=-\nabla_\mu R_{\nu\lambda} + \nabla_\lambda R_{\nu\mu} - T^{mn}{}_\mu 
     \, R_{m\nu n\lambda} + T^{mn}{}_\lambda 
     \, R_{m\nu n\mu} - T^{m}{}_{\mu\lambda} \, R_{\nu m}\;,
}
where we remind the reader that all quantities involve the torsion-full connection.
Next, we use the first Bianchi identity \eqref{bianchi_tor_01} to reorder the indices in
the two terms $T^{mn}{}_\mu  \, R_{m\nu n\lambda}$ and 
$T^{mn}{}_\lambda \, R_{m\nu n\mu}$. For the first of these expressions, this reads
\eq{
  \label{appletwo}
  T_\mu{}^{mn} R_{m\nu n\lambda} =  + \frac{1}{2} T_\mu{}^{mn} R_{mn\nu\lambda} 
  &+\frac{1}{4} T_{\mu}{}^{mn} \bigl[ \mathring \nabla_\nu T_{mn\lambda} + 2 \mathring\nabla_n T_{m\lambda\nu}
  + 3 \mathring \nabla_\lambda T_{m\nu n} \bigr] \\
  &+\frac{1}{8} T_\mu{}^{mn} \bigl[ T_{[\underline \nu \underline m}{}^p T_{\underline \lambda] np } \bigr] \;,
 }  
and similarly for the second one. 
Together with  $T^{mn}{}_\nu\, {R}_{mn\mu\lambda}$ in the first line of \eqref{variatgaussbonnet},
we then have three terms of the form
\eq{ 
\label{TR-terms}
  -T^{mn}{}_\mu 
     \, R_{mn\nu\lambda} - T^{mn}{}_\nu\, {R}_{mn\lambda\mu} -T^{mn}{}_\lambda 
     \, R_{mn\mu\nu} \;,
}
which need to be cancelled by a deformation of the 
Gauss-Bonnet action.    
This suggests the $\eta$-dependent correction 
to the curvature tensor to be
\eq{
    {\cal R}^a{}_{bcd}=
    R^a{}_{bcd}(\Gamma) +(\Delta R)^a{}_{bcd} \;,
}
where the deformation reads
\eq{
\label{curvcorrect}
   (\Delta R)^a{}_{bcd}=&-C\, \eta_{cd}{}^m\,
 \left(  T^a{}_{mb}+T_b{}^a{}_{m}+ T_{mb}{}^a{}
    \right) - 6C^2\, \eta_{cd}{}^m\, \eta^a{}_{b m}\, .
}
Note that the sum over the three torsion terms  in \eqref{curvcorrect} is not the contorsion, 
as the sign of the second term is different. Furthermore, we again included
a correction of second order  in $\eta$ so that on-shell we obtain
the Gauss-Bonnet action with the torsional connection \eqref{torsionalconn}.
Considering then terms in \eqref{variatgaussbonnet} 
which are linear both in torsion and Ricci tensors, we realize
that it is necessary to define the deformed Ricci-tensor and scalar as
\eq{
  \label{applefour}
    &{\cal R}_{ab}=R_{ab}(\Gamma) + \frac{1}{2}(\Delta
    R)_{ab} \;,\hspace{60pt}
    \widetilde{\cal R}_{ab}=\widetilde R_{ab}(\Gamma) + \frac{1}{2}(
   {\Delta \widetilde R})_{ab} \;, \\
    &{\cal R} =R(\Gamma) + \frac{1}{3}(\Delta R) \;,
}
where again $R_{ab}=R^m{}_{amb}$ and
$\widetilde R_{ab}=g^{mn}\,R_{a m n b}$.
Note that we introduced relative factors of $1/2$ and
$1/3$ between the curvature tensor and the deformation by hand.

We now consider a generalized second-order Lovelock action where the above-mentioned
deformation is included. In analogy to \eqref{action_gb}, we write
\eq{
  \label{lovelockdef}
   \mathcal S= \frac{\alpha'}{2\kappa^2}\int\! d^n x \sqrt{-g} \Bigl( {\cal R}_{abcd}\, {\cal R}^{cdab} - 
          {\cal R}_{ab}\, {\cal R}^{ba} + 2  
          {\cal R}_{ab}\, \widetilde{{\cal R}}^{ba} - 
          \widetilde{{\cal R}}_{ab}\, \widetilde{{\cal R}}^{ba} + {\cal R}^2\,
          \Bigr) \, ,
}
and for the variation with respect to $\Gamma^{\mu}{}_{\nu\lambda}$ we again use 
the notation introduced in equation \eqref{def_variation}. The contribution 
of the various terms in  \eqref{lovelockdef} then reads
\eq{
   \label{applethree}
  \renewcommand{\arraystretch}{1.75}
  \begin{array}{lcl}
   +\mathcal {R}_{abcd}\, \mathcal {R}^{cdab} &:& 
   \displaystyle \Bigl[\nabla^m( \mathcal {R}_{\nu m\mu\lambda}-\mathcal {R}_{m\nu\mu\lambda}) + T^{mn}{}_\nu\,
   \mathcal {R}_{mn\mu\lambda} + 2C\, \eta^{mn}{}_\mu \, {\cal R}_{mn\nu\lambda} \\
    &&\hspace{5pt}+ 2C\, \eta^{mn}{}_\nu\, {\cal R}_{mn\lambda\mu} +2C\, \eta^{mn}{}_\lambda 
     \, {\cal R}_{mn\mu\nu}
    \Bigr]  \;, \\
   -\mathcal {R}_{ab}\, \mathcal {R}^{ba} &:& \displaystyle \Bigl[ \nabla_\mu 
   \mathcal {R}_{\nu\lambda} - g_{\mu\nu}\,
    \nabla^m \mathcal {R}_{m\lambda} - T_{\nu\mu}{}^m\, \mathcal {R}_{m\lambda} 
    -C\, \eta_{\mu\nu}{}^m\,
    {\cal R}_{m\lambda} \\
   &&\hspace{5pt}-C\, \eta_{\nu\lambda}{}^m\, {\cal R}_{m\mu}  -C\, \eta_{\lambda\mu}{}^m\, {\cal R}_{m\nu} 
    \Bigr] \;, \\
   -\widetilde{\mathcal R}_{ab}\, \widetilde{\mathcal R}^{ba} 
   &:& \displaystyle\Bigl[\nabla_\lambda \widetilde{\mathcal R}_{\nu\mu} - g_{\nu\lambda}\,
    \nabla^m \widetilde{\mathcal R}_{m\mu} - T_{\nu\lambda}{}^m\, \widetilde{\mathcal R}_{m\mu} 
    +C\, \eta_{\mu\nu}{}^m\,
    \widetilde{\cal R}_{m\lambda} \\
   &&\hspace{5pt}+C\, \eta_{\nu\lambda}{}^m\, \widetilde{\cal R}_{m\mu}  +C\, \eta_{\lambda\mu}{}^m\, 
   \widetilde{\cal R}_{m\nu}     \Bigr] \;, \\
   +2{\mathcal R}_{ab}\, \widetilde{\mathcal R}^{ba} 
   &:& \displaystyle \Bigl[-\nabla_\mu \widetilde{\mathcal R}_{\nu\lambda} + g_{\mu\nu}\,
    \nabla^m \widetilde{\mathcal R}_{m\lambda} + T_{\nu\mu}{}^m\, \widetilde{\mathcal R}_{m\lambda}
    +C\, \eta_{\mu\nu}{}^m\,
    \widetilde{\cal R}_{m\lambda} \\
   &&\hspace{5pt}+C\, \eta_{\nu\lambda}{}^m\, \widetilde{\cal R}_{m\mu}  +C\, \eta_{\lambda\mu}{}^m\, 
   \widetilde{\cal R}_{m\nu}    \\
   && \displaystyle \hspace{5pt}-\nabla_\lambda {\mathcal R}_{\nu\mu} +g_{\nu\lambda}\,
    \nabla^m {\mathcal R}_{m\mu} + T_{\nu\lambda}{}^m\, {\mathcal R}_{m\mu}
    -C\, \eta_{\mu\nu}{}^m\,
    {\cal R}_{m\lambda} \\
   &&\hspace{5pt}-C\, \eta_{\nu\lambda}{}^m\, {\cal R}_{m\mu}  -C\, \eta_{\lambda\mu}{}^m\, {\cal R}_{m\nu}     
    \Bigr]  \;, \\
  +\mathcal {R}^2 &:& 
  \displaystyle \Bigl[ g_{\mu\nu}\, \nabla_\lambda \mathcal {R}- g_{\nu\lambda}\, \nabla_\mu \mathcal{R} 
  +T_{\nu\mu\lambda}\, \mathcal {R} +2 C \eta_{\mu\nu\lambda} \mathcal R\Bigr]   
  \;,
   \end{array}
}
and the equation of motion for the connection is again the sum of these terms being equal to zero.
To make contact with the solution at linear order in curvature, we require that
the field equation for $\Gamma$ is satisfied 
for $T_{abc}=2C\, \eta_{abc}$. One could in principle allow for 
 $\alpha'$  corrections to this relation, but this would induce
corrections at order $(\alpha')^2$ in the equation of motion, which
could then only be cancelled by including additional  higher order terms. 
We want to avoid such tuning between different terms in
the $\alpha'$ expansion,  so that we require that
the relation $T_{abc}=2C\, \eta_{abc}$ holds at each order separately.

Since in this case  the connection is metric compatible, we can employ the relations
\eq{    
{\cal R}_{abcd}=-{\cal R}_{bacd}\; , \hspace{60pt}
\widetilde {\cal R}_{ab}=-{\cal R}_{ab} \;.
}
In addition to the relations \eqref{appleone} and \eqref{appletwo},  we contract indices in
\eqref{appleone} to obtain
\eq{
     \nabla^m R_{mc}= \frac{1}{ 2} \nabla_c R + T_c{}^{mn}\, R_{mn} 
             +\frac{1}{ 2} T^{mnp}\, R_{mnpc}\;,
}
which we use to cancel all terms of the form
$\nabla_{\star} R_{\star\ldots \star}$
and $\eta^{\star\star\star} R_{\star\ldots \star}$ in \eqref{applethree}.
In fact, the deformations to the curvature tensors in \eqref{curvcorrect} and \eqref{applefour}
are designed to precisely achieve this.
The only remaining terms in the field equation for the connection $\Gamma$ are of the
schematic form
$\nabla (\eta^{\star\star\star} \eta_{\star\star\star})$ and $(\eta_{\star\star\star})^3$,
which can be summarized as
\eq{ 
\label{finalrels1}
  F_{\mu\nu\lambda} = +2C^2 \Bigl[ &+ \eta_\mu{}^{mn} \bigl( -\mathring\nabla_\nu \eta_{\lambda mn}
    + 2 \mathring\nabla_n \eta_{m\nu\lambda} + 3 \mathring\nabla_\lambda \eta_{\nu mn} \bigr) \\
    &- \eta_\lambda{}^{mn} \bigl( -\mathring\nabla_\nu \eta_{\mu mn}
    + 2 \mathring\nabla_n \eta_{m\nu\mu} + 3 \mathring\nabla_\mu \eta_{\nu mn} \bigr) \\
    &+g_{\mu\nu} \bigl( 2 \eta_\lambda{}^{mn} \mathring\nabla^p \eta_{pmn}  + \eta^{pmn} 
    \mathring\nabla_p \eta_{mn\lambda} - \eta^{mnp} \mathring\nabla_\lambda \eta_{mnp} \bigr) \\
    &-g_{\nu\lambda} \bigl( 2 \eta_\mu{}^{mn} \mathring\nabla^p \eta_{pmn}  + \eta^{pmn} 
    \mathring\nabla_p \eta_{mn\mu} - \eta^{mnp} \mathring\nabla_\mu \eta_{mnp} \bigr) \\
    &+4 \eta_{\mu\lambda}{}^m \mathring \nabla^p \eta_{p\nu m}
    \quad\Bigr] \\
    + 2 C^3 \Bigl[ & + \eta_\mu{}^{mn} \bigl( \eta_{m[\underline n}{}^p \eta_{\underline \nu \underline \lambda]p}
    \bigr)
    - \eta_\lambda{}^{mn} \bigl( \eta_{m[\underline n}{}^p \eta_{\underline \nu \underline \mu]p}
    \bigr) \quad \Bigr] \;.
}
Since this expression is neither symmetric nor cyclic in the 
three indices $\{\mu,\nu,\lambda\}$, it seems hopeless to find a diffeomorphism
invariant additional deformation to the curvature tensors  to cancel it.
Therefore, we suspect that  \eqref{finalrels1} must vanish.
Note that we have not yet employed a Bianchi identity
for the three-index object $\eta_{abc}$. Thus, one might suspect that
for instance the Bianchi identity for $H$-flux, that is $dH=0$, or in components
\eq{
  \label{bianchi_h}
  0 = \mathring\nabla_{[\underline a} \eta_{\underline{bcd}]}  \;,
}
makes $F_{\mu\nu\lambda}$ vanishing. However, using \eqref{bianchi_h} we can rewrite \eqref{finalrels1} as
\eq{
\label{finalrels}
  F_{\mu\nu\lambda} = +4C^2 \Bigl[ &+ \eta_\mu{}^{mn}  \mathring\nabla_\lambda \eta_{\nu mn}  
    - \eta_\lambda{}^{mn}  \mathring\nabla_\mu \eta_{\nu mn}  +2 \eta_{\mu\lambda}{}^m \mathring \nabla^p \eta_{p\nu m} \\
    &+g_{\mu\nu} \bigl(  \eta_\lambda{}^{mn} \mathring\nabla^p \eta_{pmn}  -{\textstyle \frac{1}{3}}\, \eta^{mnp} 
    \mathring\nabla_\lambda \eta_{mnp} \bigr) \\
    &-g_{\nu\lambda} \bigl(  \eta_\mu{}^{mn} \mathring\nabla^p \eta_{pmn}   
    -{\textstyle \frac{1}{3}}\, \eta^{mnp} 
    \mathring\nabla_\mu \eta_{mnp} \bigr) \quad \Bigr] \\
    + 2 C^3 \Bigl[ & + \eta_\mu{}^{mn} \bigl( \eta_{m[\underline n}{}^p \eta_{\underline \nu \underline \lambda]p}
    \bigr)
    - \eta_\lambda{}^{mn} \bigl( \eta_{m[\underline n}{}^p \eta_{\underline \nu \underline \mu]p}
    \bigr) \quad \Bigr] \;,
}
and we see that $F_{\mu\nu\lambda}$ does not vanish.
The  only reasonable and general (i.e. Bianchi-type) conditions 
to guarantee $F_{\mu\nu\lambda}=0$ are 
\eq{\label{jacobi}
\bom{
 \nabla_a \eta_{bcd}=0\; , 
       \hspace{60pt}
       \eta_{[\underline a}{}^{mn}\, \eta_{\underline b\underline c]\, n}=0 \;,
}
}        
that is the three-index object is covariantly constant and satisfies 
a  Jacobi identity. Note furthermore,  due to the latter relation
we have $\nabla_a \eta_{bcd}=\mathring\nabla_a \eta_{bcd}=0$.
In section \ref{sec:bianchi} we will point out and elaborate on a possible origin of the 
relations  \eqref{jacobi} in the context of string theory.
Inserting then the connection into the deformed Gauss-Bonnet action
\eqref{lovelockdef}, we obtain the usual Gauss-Bonnet action
with torsion
\eq{\label{love_2}
    \mathcal S=\frac{\alpha'}{2\kappa^2}
  \int d^n x \sqrt{-g}\, \Bigl[ {R}(\Gamma)_{abcd}\, {R}(\Gamma)^{cdab} - 
          4{R}(\Gamma)_{ab}\, {R}(\Gamma)^{ba} + {R}(\Gamma)^2\, \Bigr] \; .
}
A quite tedious computation shows that this action is indeed
a consistent truncation of \eqref{lovelockdef}  with
respect to variation of both the metric and the Kalb-Ramond field.
We emphasize  that this is not a trivial result and a number
of cancellations occur in the course of the computation.

To conclude, we want to stress  that we  introduced the deformations \eqref{curvcorrect} and \eqref{applefour}  in order to cancel the terms \eqref{TR-terms}. Moreover, the emergence of the conditions \eqref{jacobi}  to solve the equations of motion becomes conspicuous in this approach. However, the deformations are not unique, and in the following section we  show that there exists a different and more elegant  deformation.


\section{Palatini-Lovelock-Cartan actions}
\label{sec:lovelock}

In this section, we show that taking the two  conditions \eqref{jacobi}
on the three-form into account from the very beginning, an even
simpler Palatini-type Gauss-Bonnet action can be found. And, as we show, this structure generalizes to higher-order Lovelock actions \cite{Lovelock:1971yv}.


\subsubsection*{The Lovelock action}

Let us recall 
a  result of Exirifard and Sheikh-Jabbari \cite{Exirifard:2007da}. 
These authors have studied  higher-curvature actions
which are consistent with the Palatini variational principle, and have found
that precisely for Lovelock actions  the
$\Gamma$ equation of motion is consistently
satisfied by the Levi-Civita connection. Also, requiring a consistent truncation for the Palatini approach singles-out these actions \cite{Dadhich:2010dg}.

The Lovelock action at order $k$ in the curvature tensor can be expressed in the following way
\eq{
  \label{lovelock}
   S^{(k)}_{\rm Love} &=
   \frac{(\alpha')^{k-1}}{4\kappa^2 k!} \int \star \left(e^{a_1}\wedge e^{b_1}\wedge\ldots\wedge e^{a_k}\wedge e^{b_k}
    \right) \wedge  \left( \Omega_{a_1 b_1}\wedge\ldots\wedge \Omega_{a_k b_k}\right) \\
	&= \frac{(\alpha')^{k-1}}{4\kappa^2k!}\int d^n x \,\sqrt{-g}\,  
  \left\vert \begin{matrix}  g^{a_1 b_1} & \ldots & g^{a_1 b_{2k}} \\
               \vdots &  & \vdots \\
    g^{a_{2k} b_1} & \ldots & g^{a_{2k} b_{2k}} 
                    \end{matrix}\right\vert\, 
    \prod_{j=1}^k R_{a_{2j-1}a_{2j}b_{2j-1}b_{2j}}\;,
}
where $\sqrt{\alpha'}$ is a fundamental length scale. 
Here, $\{e^a\}$ denotes a set of vielbeins and 
 $\Omega_{a b}$ is the curvature two-form
\eq{
      \Omega^a{}_b= \frac{1}{ 2}\, R^a{}_{bcd}\,e^c\wedge e^d =d\omega^a{}_b+\omega^a{}_c\wedge \omega^c{}_b \;,
}
with $\omega^a{}_b=\Gamma^a{}_{cb}\, e^c$ being  the connection one-form. For $k=1$ we obtain from \eqref{lovelock} the Einstein-Palatini action \eqref{actionEP}, and for $k=2$ the Palatini-Gauss-Bonnet action \eqref{action_gb}.

Assuming the constraints \eqref{jacobi} to be valid, we now show that the deformation of the curvature tensor employed already for the Einstein-Palatini case \eqref{actionEPtors} 
is also a viable solution at  higher orders. In particular, we define a deformed Riemann curvature tensor as
\eq{
  \label{deformation_new}
  \mathcal R^{a}{}_{bcd} = 
  R^{a}{}_{bcd}(\Gamma) + C \eta^{a}{}_{b m} T^m{}_{cd}
  - 2 C^2 \eta^{a}{}_{b m} \eta^m{}_{cd} \;,
}
and do not  impose relative factors between the curvature tensors 
and the deformation as we did in section \ref{sec:GB}.
Since the case $k=1$ was considered already in section \ref{sec_linear}, we consider  the case $k=2$.


\subsubsection*{The second-order case}

The Lovelock action at second order in the curvature can be obtained from \eqref{lovelock} and reads
\eq{
  \label{action_gb_05}
   \mathcal S^{(2)}_{\rm Love}= \frac{\alpha'}{2\kappa^2}\int d^n x \sqrt{-g}\, \Bigl( {\cal R}_{abcd}\, {\cal R}^{cdab} - 
          &{\cal R}_{ab}\, {\cal R}^{ba} \\[-0.1cm]
    &+ 2  
          {\cal R}_{ab}\, \widetilde{{\cal R}}^{ba} - 
          \widetilde{{\cal R}}_{ab}\, \widetilde{{\cal R}}^{ba} + {\cal R}^2\,
          \Bigr) ,
}
where again $\mathcal R_{ab} = \mathcal R^m{}_{amb}$, $\widetilde{\mathcal R}_{ab} = \mathcal R_{a}{}^m{}_{mb}$ and $\mathcal R = \mathcal R^{ab}{}_{ab}$.
Employing then the notation introduced in \eqref{def_variation}, the variation of the action \eqref{action_gb_05} with respect to the connection can be written as 
\eq{
  \label{variation_17}
  F_{\mu\nu\lambda} = &- \nabla^m \mathcal R_{m\nu\mu\lambda} +  \nabla^m \mathcal R_{\nu m\mu \lambda} + T_{\nu}{}^{mn}
  \mathcal R_{mn\mu\lambda}
  + C  \eta_{\mu}{}^{mn} (\mathcal R_{\nu\lambda mn} - \mathcal R_{\lambda\nu mn} )
  \\[2mm]
  &+\nabla_\mu ( \mathcal R_{\nu\lambda} - \widetilde{\mathcal R}_{\nu\lambda} ) 
  -  \nabla_\lambda  ( \mathcal R_{\nu\mu} - \widetilde{\mathcal R}_{\nu\mu} ) \\
  &
  - g_{\mu\nu}\nabla^m  ( \mathcal R_{m\lambda} - \widetilde{\mathcal R}_{m\lambda} )
  + g_{\nu\lambda}\nabla^m  ( \mathcal R_{m\mu} - \widetilde{\mathcal R}_{m\mu} )
  \\
  & + T_{\nu\mu}{}^m  ( \mathcal R_{m\lambda} - \widetilde{\mathcal R}_{m\lambda} ) 
  - T_{\nu\lambda}{}^m  ( \mathcal R_{m\mu} - \widetilde{\mathcal R}_{m\mu} ) \\
  & + 2 C \left[ \eta_{\mu\nu}{}^m  ( \mathcal R_{\lambda m} - \widetilde{\mathcal R}_{\lambda m} ) 
  + \eta_{\lambda\mu}{}^m  ( \mathcal R_{\nu m} - \widetilde{\mathcal R}_{\nu m} ) \right]\\[2mm]
  &+ g_{\mu\nu} \nabla_\lambda \mathcal R - g_{\nu\lambda} \nabla_\mu \mathcal R + \left( T_{\nu\mu\lambda} + 2 C \eta_{\mu\nu\lambda} \right) \mathcal R \;.
}

Next, similarly as  for the analysis at  linear order in the curvature, we require that the field equation for the connection is solved by a metric-compatible connection with  $T_{abc}=2C\, \eta_{abc}$. In this case, we can replace $\mathcal R_{\star \ldots \star}$ by $R_{\star \ldots \star}$ in \eqref{variation_17}, as can be seen from  \eqref{deformation_new}, leading to
\eq{
  \label{variation_18}
  F_{\mu\nu\lambda} = &+2 \nabla^m  R_{\nu m\mu\lambda} + 2 \nabla_\mu  R_{\nu\lambda} - 2 \nabla_\lambda  R_{\nu\mu} \\
  & - g_{\mu\nu} \bigl(\, 2 \nabla^m  R_{m\lambda} - \nabla_\lambda  R \bigr)
  + g_{\nu\lambda} \bigl(\, 2 \nabla^m  R_{m\mu} - \nabla_\mu  R \bigr) \\
  &+ 4C \eta_{\mu\lambda}{}^m  R_{m\nu} + 4C \eta_{\nu\lambda}{}^m  R_{m\mu}
  + 2 C \eta_{\nu}{}^{mn}    R_{mn\mu\lambda}  + 2 C \eta_{\mu}{}^{mn}    R_{mn\nu\lambda} \;.
}
Furthermore, making use of the first constraint in \eqref{jacobi}, from $[\nabla_m,\nabla_n ] \eta_{abc}=0$ we can derive
\eq{
  \label{new_rel_18}
   0 = 
     R^p{}_{amn} \eta_{pbc} 
   +R^p{}_{bmn} \eta_{apc}
   +R^p{}_{cmn} \eta_{abp} \;.
}
Employing then \eqref{jacobi} as well as \eqref{new_rel_18}, we observe that the Bianchi identities shown in \eqref{bianchi_tor_01} and \eqref{bianchi_tor_02} simplify to 
\eq{
   R^a{}_{[\underline{bcd}]} =   0 \;, 
   \hspace{50pt}
     \nabla_{[\underline{a}} R^{mn}{}_{\underline{bc}]}=0 \;.
}
Finally, using the second of these equations we see that the first two lines in \eqref{variation_18} vanish, and using the first Bianchi identity  together with \eqref{new_rel_18}, the third line
in  \eqref{variation_18} vanishes. We therefore arrive at
\eq{
  F_{\mu\nu\lambda} = 0\;,
}
that is the variation of the action \eqref{action_gb_05} with respect to the connection vanishes for a metric-compatible connection with  torsion $T_{abc}=2C\, \eta_{abc}$.
Furthermore, the same computation can be performed with a more general
deformed curvature tensor of the form
\eq{
  \mathcal R^{a}{}_{bcd} = 
  R^{a}{}_{bcd}(\Gamma) + C \eta^{a}{}_{b m} T^m{}_{cd}
  - (2 C^2-A)\, \eta^{a}{}_{b m} \eta^m{}_{cd} \;.
}
In this case, consistency with the leading order string action \eqref{stringaction} yields 
the relation $C^2-A=1/12$.


\subsubsection*{The third-order case}

After having studied the second order Lovelock action, we now consider the case of $k=3$. From \eqref{lovelock} we find the third-order Lovelock action as \cite{MuellerHoissen:1985mm,Wheeler:1985qd}
\begin{align}
  \label{action_love_01}
   \mathcal S^{(3)}_{\rm Love}&= \frac{(\alpha')^2}{2\kappa^2}\,\frac{2}{3}\int d^n x \sqrt{-g}\, \Bigl[
     \;{\cal R}^{abcd}\, {\cal R}_{cdef}\, {\cal R}^{ef}{}_{ab} +4\, 
     {\cal R}^{ab}{}_{ce}\,  {\cal R}^{cd}{}_{bf}\, {\cal
       R}^{ef}{}_{ad}\nonumber \\[0.1cm]
    &\qquad\qquad +3\, {\cal R}^{abcd}\, {\cal R}_{cdbe}\, {\cal R}^e{}_{a}
     -3\, {\cal R}^{abcd}\, {\cal R}_{cdae}\, {\cal R}^e{}_{b}\nonumber \\[0.1cm]
    &\qquad\qquad -3\, {\cal R}^{abcd}\, {\cal R}_{cdbe}\, \widetilde{\cal R}^e{}_{a}
    +3\, {\cal R}^{abcd}\, {\cal R}_{cdae}\, \widetilde{\cal R}^e{}_{b}
    \\[0.1cm]
    &\qquad\qquad +3\, {\cal R}^{acbd}\, {\cal R}_{ba}\, {\cal R}_{dc}
    -6\, {\cal R}^{acbd}\, {\cal R}_{ba}\, \widetilde{\cal R}_{dc}
    +3\, {\cal R}^{acbd}\, \widetilde{\cal R}_{ba}\, \widetilde{\cal
      R}_{dc}\nonumber \\[0.1cm]
    &\qquad\qquad + {\cal R}^{ab}\, {\cal R}_{bc}\,  {\cal R}^c{}_{a}
    - 3\, {\cal R}^{ab}\, {\cal R}_{bc}\,  \widetilde{\cal R}^c{}_{a}
    + 3\, {\cal R}^{ab}\, \widetilde{\cal R}_{bc}\,  \widetilde{\cal R}^c{}_{a}
    -  \widetilde{\cal R}^{ab}\, \widetilde{\cal R}_{bc}\,  \widetilde{\cal
      R}^c{}_{a}\nonumber \\
    & \qquad\qquad+ {\cal R} \times (\mbox{ lower order Lovelock terms}) \quad
    \Bigr] \;. \nonumber
\end{align}
Employing again our previous notation, the variation of the above action 
with respect to the connection leads to
\eq{
  \label{tedious_01}
   F_{\mu\nu\lambda}= \frac{4}{3}\,C \Bigl[ &+3\,\eta^{mn}{}_\nu\, R_{mnpq}\, R^{pq}{}_{\mu\lambda} 
         +12\, \eta^{mn}{}_\nu\, R_{mp\lambda q}\, R_{n}{}^{q}{}_{\mu}{}^p\\[-4pt]
         &+6\, \eta^{mn}{}_\nu\, R_{mn\lambda p}\, R^p{}_\mu 
           -6\, \eta^{mn}{}_\nu\, R_{mn\mu p}\, R^p{}_\lambda \\
         &+12\, \eta^{mn}{}_\nu\, R_{pm\mu\lambda}\, R^p{}_n
         +12\,  \eta^{mn}{}_\nu\, R_{m\mu}\, R_{n\lambda}\\
         &-6 \,   \eta_{\mu\nu}{}^m\, R_m{}^{npq}\, R_{pqn\lambda}
         -6 \,   \eta_{\nu\lambda}{}^m\, R_m{}^{npq}\, R_{pqn\mu}\\
          &-12\, \eta_{\mu\nu}{}^m\, R_{mp\lambda q}\, R^{pq}
          -12\, \eta_{\nu\lambda}{}^m\, R_{mp\mu q}\, R^{pq}\\[-4pt]
         &-12\, \eta_{\mu\nu}{}^m\, R_{mp}\, R^{p}{}_\lambda
         -12\, \eta_{\nu\lambda}{}^m\, R_{mp}\, R^{p}{}_\mu \qquad\Bigr] 
	+ \bigl(\mu\leftrightarrow \nu\bigr) \;,
}
where the terms arising from the last line in \eqref{action_love_01}
cancel due to our analysis for the lower-order Lovelock actions.
After a tedious calculation and
employing the following four relations descending from \eqref{new_rel_18}
\begin{align}
\eta^{mn}{}_b\, R_{mnap}\,  R^p{}_c +\eta^{nmp}\, R_{pmab}\,  R_{nc}&=
       -\eta^{mn}{}_b\, R_{mncp}\,  R^p{}_a -\eta^{nmp}\, R_{pmcb}\,  R_{na}
       \;,
\nonumber\\[0.2cm]
     \eta^{mn}{}_b\, R_{amcp}\,  R^p{}_n +\eta^{mnq}\, R_{mpbn}\,  R_{qa}{}^p{}_c&=
      -\eta^{mn}{}_b\, R_{cmap}\,  R^p{}_n -\eta^{mnq}\, R_{mpbn}\,
      R_{qc}{}^p{}_a \;, \nonumber \\[0.2cm]
   \eta^{nmp}\, R_{pmab}\, R_{nc}&=-2\, \eta^{mnq}\, R_{mpcn}\,  R_{qb}{}^p{}_a 
    +2\, \eta^{mnq}\, R_{mpcn}\,  R_{qa}{}^p{}_b \;, \nonumber\\[0.2cm]
   \eta^{mnq}\, R_{mpan}\,  R_{qb}{}^p{}_c &=
  - \eta^{mnq}\, R_{mpcn}\,  R_{qb}{}^p{}_a \;,
\end{align}
we can bring \eqref{tedious_01} into the  form
\eq{
   F_{\mu\nu\lambda}=\frac{4}{3}\,C \biggl[ 
   & +6\,\Bigl( \eta^{mn}{}_\nu\, R_{mn\mu p}\,  R^p{}_\lambda -
                   \eta^{mn}{}_\mu\, R_{mn\nu p}\,  R^p{}_\lambda \Bigr)\\[-3pt]
         &+3\,\Bigl( \eta^{mn}{}_\lambda\, R_{mn\nu p}\,  R^p{}_\mu -
                   \eta^{mn}{}_\lambda\, R_{mn\mu p}\,  R^p{}_\nu \Bigr)\\
       &+12\,\Bigl( \eta^{mn}{}_\nu\, R_{\mu m\lambda p}\,  R^p{}_n -
                  \eta^{mn}{}_\mu\, R_{\nu m\lambda p}\,  R^p{}_n  \Bigr)\\
       &+6\,\Bigl( \eta^{mn}{}_\lambda\, R_{\nu m\mu p}\,  R^p{}_n -
                  \eta^{mn}{}_\lambda\, R_{\mu m\nu p}\,  R^p{}_n  \Bigr)\\
        &+6\,\Bigl( \eta^{mn}{}_\nu\, R_{m\lambda pq}\,  R_{n\mu}{}^{pq} -
                  \eta^{mn}{}_\mu\, R_{m\lambda pq}\,  R_{n\nu}{}^{pq} \Bigr)\\[-3pt]
        &+6 \,  \eta^{mn}{}_\lambda\, R_{m\mu pq}\,  R_{n\nu}{}^{pq}
        \hspace{120pt} \biggr] + \bigl( \mu\leftrightarrow \nu\bigr) \;.
}
Since all terms written out explicitly are anti-symmetric
under the exchange $\mu\leftrightarrow \nu$, we conclude that
\eq{
  F_{\mu\nu\lambda} = 0 \;.
}  
Therefore, also for the third-order  Palatini-Lovelock-Cartan action 
the torsion-full connection with $T^a{}_{bc} = 2 C \eta^a{}_{bc}$ is a
solution to the field equation of $\Gamma$.
Let us emphasize that, at least to us,  this result is non-trivial.


\subsubsection*{A conjecture for Palatini-Lovelock-Cartan gravity}

The above observations and results lead us to the following conjecture:

\begin{itemize}

\item[]{\bf Conjecture:} {\it The field equation of the connection of the
Palatini-Lovelock-Cartan action}
\eq{  
  \label{love_def}
  \mathcal S^{(k)}_{\rm Love}
  &=\frac{(\alpha')^{k-1}}{4\kappa^2k!}\int d^n x \,\sqrt{-g}\,  
  \left\vert \begin{matrix}  g^{a_1 b_1} & \ldots & g^{a_1 b_{2k}} \\
               \vdots &  & \vdots \\
    g^{a_{2k} b_1} & \ldots & g^{a_{2k} b_{2k}} 
                    \end{matrix}\right\vert\, 
    \prod_{j=1}^k {\cal R}_{a_{2j-1}a_{2j}b_{2j-1}b_{2j}}
\nonumber
} 
{\it with the deformed curvature tensor}
\eq{
   \mathcal R^{a}{}_{bcd} = 
  R^{a}{}_{bcd}(\Gamma) + C \eta^{a}{}_{b m} T^m{}_{cd}
  - 2 C^2 \eta^{a}{}_{b m} \eta^m{}_{cd} \;\nonumber
}         
{\it is satisfied for a metric-compatible connection with $T_{abc}=2\,C\eta_{abc}$, 
if the three-form $\eta_{abc}$ is covariantly constant
and satisfies the Jacobi identity \linebreak $\eta_{[\underline a}{}^{mn}\, \eta_{\underline b \underline c]\, n}=0$. Inserting the
solution into the Lovelock-Cartan action results in a consistent truncation.}
\end{itemize}


\section{Relation to bosonic string theory}
\label{sec:ST}

Our analysis  so far was based on a rather  formal question, namely
whether the Palatini variational principle can be generalized 
to torsional Gauss-Bonnet gra\-vi\-ty. 
In this section, we study   
whether the Gauss-Bonnet action with torsion
can be related to  the effective action at second order in curvature
of the bosonic string for constant dilaton.
This was already analyzed a long time ago in \cite{Bern:1987wz} 
with a negative answer. However, in view of 
the new  Bianchi-type identities shown in \eqref{jacobi} it seems worthwhile to redo 
their analysis.

Let us recall parts of the results of \cite{Bern:1987wz}.
Using equation \eqref{curvexpand} and invoking  the Bianchi identities for
the curvature tensor with Levi-Civita  connection as well as for the $H$-flux \eqref{bianchi_h}, 
it was shown that any action quadratic in the torsional 
curvature can be expressed by  ten independent combinations
\eq{  \mathcal S= \frac{\alpha'}{2\kappa^2}
   \int d^n x \sqrt{-g}\; \Bigl[ \sum_{i=1}^3 f_i\; T_{f_i}
     + \sum_{j=1}^4 a_j\; T_{a_j} + \sum_{k=1}^3 u_k\; T_{u_k}
\Bigr] \;.
}
The above terms read
\eq{
  \renewcommand{\arraystretch}{1.2}
  \arraycolsep2pt
  \begin{array}[t]{lcl}
   T_{f_1}&=&\mathring {R}_{abcd}\, \mathring{R}^{abcd}\;, \\
   T_{f_2}&=&\mathring{R}_{abcd}\, \eta^{ac m}\, \eta^{bd}{}_m\;, \\
   T_{f_3}&= &\eta_{abc}\, \eta^{a m}{}_n\, \eta^{b}{}_{mp}\, \eta^{c np}\;, \\[0.3cm]
    T_{u_1}&=& \mathring{R}_{ab} \mathring{R}^{ab} \;,\\
    T_{u_2}&= &\mathring{R}^2 \;,\\
    T_{u_3}&=& (\mathring\nabla_m\, \eta_{ab}{}^m )\, (\mathring\nabla_m\, \eta^{ab m} )\; ,
    \end{array}
    \hspace{40pt}
  \begin{array}[t]{lcl}
   T_{a_1}&=& (\eta_{a mn}\, \eta_{b}{}^{mn}) (\eta^{a pq}\, \eta^{b}{}_{pq})\;, \\ 
   T_{a_2}&=& (\eta_{abc}\, \eta^{abc})^2\;, \\
   T_{a_3}&=& \mathring{R}\, (\eta_{abc}\, \eta^{abc}) \;, \\ 
   T_{a_4}&= &\mathring{R}_{ab} (\eta^{a mn}\, \eta^{b}{}_{mn})\;, \\[0.2cm]
\end{array}
}
where we employed the same the notation as in \cite{Bern:1987wz}.  
As can be seen for instance from the redefinition of the metric,
the Kalb-Ramond field and the dilaton 
(see e.g.  \cite{Tseytlin:1986zz,Hull:1987yi,Jones:1988hk}), the on-shell string
scattering amplitudes impose four relations among the
coefficients of these ten terms.
They are given by
\eq{\label{strelB}
    f_1=1\; ,\hspace{30pt} 
    f_2=-1 \;,\hspace{30pt}
     f_3={\textstyle \frac{1}{ 24}}\;  ,\hspace{30pt}
    a_1+{\textstyle\frac{1}{ 4}} a_4 +{\textstyle \frac{1}{ 16}} u_1=-
   {\textstyle\frac{1}{ 8}}\; .
} 
The two conditions  in  \eqref{jacobi} allow for a reduction 
to seven terms, where  $T_{u_3}$ vanishes 
identically and where we can relate
\eq{\label{f3a1}
	2\, T_{f_3}=T_{a_1}\;, 
	\hspace{40pt} 2\, T_{f_2}=T_{a_4} \; .
}
Thus the relations \eqref{strelB} can be simplified to
\eq{\label{strel}
	f_1=1 \;,\hspace{40pt}
    a_1+{\textstyle\frac{1}{ 4}} a_4 +{\textstyle\frac{1}{ 16}} u_1=-
   {\textstyle\frac{11}{ 48}}\; .
}
Now, using \eqref{curvexpand}, we expand the curvature terms in the 
truncated action \eqref{love_2} to obtain
\eq{
   \mathcal S = \frac{\alpha'}{2\kappa^2} \int d^n x\sqrt{-g}\;\Bigl(T_{f_1}
    -3C^4\,T_{a_1} +C^4 \, T_{a_2} &-2C^2 \, T_{a_3}\\ 
    & + 6C^2\, T_{a_4} - 4 \,T_{u_1} + T_{u_2} \Bigr) \;.
}
As the second relation in \eqref{strel} is off by $1/12$,
the pure torsional Gauss-Bonnet action is not consistent
with the effective action of the bosonic string. Therefore,
even with the stronger Bianchi-type identities the
negative result of \cite{Bern:1987wz} still holds.

However,  it is possible to add an additional correction to the Gauss-Bonnet
action to make it consistent.
Defining the tensor
\eq{
      P_{abcd}=\frac{1}{ 6}\, \eta_{ab}{}^m\,  \eta_{cdm}
}
and contracting indices as for the curvature tensor, we obtain
the effective action of the bosonic string as
\eq{
    \mathcal S_{\rm string}=\frac{\alpha'}{2\kappa^2}\int d^n x \sqrt{-g} \Bigl( \hspace{10pt}
 &\Bigl[ {R}_{abcd}\, {R}^{cdab} - 
          4{R}_{ab}\, {R}^{ba} + {R}^2 \Bigr]
\\ +&\Bigl[ {P}_{abcd}\, {P}^{cdab} - 
          4{P}_{ab}\, {P}^{ba} + {P}^2 \Bigr]\hspace{10pt}\Bigr) \; ,
}
where we used the short-hand notation  ${R}_{abcd}={R}(\Gamma)_{abcd}$.
Note that this is still compatible with the Palatini formalism since we included terms only
depending on $\eta_{abc}$. 
Furthermore, as can be checked, the 
additional term already included in the off-shell action \eqref{lovelockdef}
leads to a consistent  truncation as well.


\section{Bianchi identities}
\label{sec:bianchi}

We have seen  that 
our Palatini approach to the 
Gauss-Bonnet action with torsion imposes  
conditions on the three-form flux which go beyond the expected
Bianchi identity for the $H$-flux
\eq{
\label{bianchih}
       \mathring \nabla_{[\underline a}\, H_{\underline{bcd}]}=0\, .
}
In particular, in \eqref{jacobi} we observed that  
the three-form flux should be covariantly constant and
should satisfy a Jacobi identity. 
Such conditions are not completely unfamiliar, since similar
constraints appear for certain exact solutions to
the string equations of motion, namely for parallelizable 
manifolds such as WZW models \cite{Braaten:1985is}.

Moreover, the relations we found resemble  Bianchi identities not for the $H$-flux itself 
but for the other three T-dual counterparts: geometric flux $f$
and non-geometric fluxes $Q$ and $R$.
In \cite{Ihl:2007ah} the form of these Bianchi identities was derived
as Jacobi identities of generalized gauge transformations for {\em constant}
fluxes on a flat geometry 
\eq{
\label{wrasebianchi}
H_{k[\underline{ab}}\, f^k{}_{\underline{cd}]} = 0\;, &  \\
H_{k[\underline{ab}}\, Q_{\underline{c}]}{}^{kj} + f^j{}_{k[\underline{a}}\, f^{k}{}_{\underline{bc}]} =0\;,&  \\
H_{kab}\, R^{kcd} + f^k{}_{ab}\, Q_k{}^{cd} -
4f^{[\underline{c}}{}_{k[\underline{a}} 
\, Q_{\underline{b}]}{}^{\underline{d}]k} =0\;, &  \\
f^{[\underline{a}}{}_{ki}\, R^{\underline{bc}]k} + Q_i{}^{k[\underline{a}}\,Q_k{}^{\underline{bc}]} =0\;,&  \\
Q_k{}^{[\underline{ab}}\, R^{\underline{cd}]k} =0 \; .&
}
Note that due to their origin, the left hand side of all these relations can be written
as sums over terms of the form $\eta_{[\underline a}{}^{mn}\, \eta_{\underline b\underline c]\, n}$.

For non-constant fluxes on general manifolds one 
expects corrections to the above
Bianchi identities containing derivative and curvature terms. 
It is the purpose of this section to derive  
these corrections which to large extend are unknown.


\subsubsection*{Bianchi identity for $Q$- and $R$-flux}

Let us  propose a possible way to   derive the Bianchi identities for 
the non-geometric $Q$- and $R$-flux, which should be related
to the last three relations in \eqref{wrasebianchi}. 
The $R$-flux can be described  as an anti-symmetric three-vectorfield,
and so we expect defining
relations to be determined by the use of
a graded extension of the Lie bracket of vectorfields. 

For this purpose we introduce the \emph{Schouten--Nijenhuis bracket} $[\ ,\ ]_{SN}$, 
which for functions $f,g$ and ordinary vectorfields $X,Y$ it is defined by
\eq{
[f,g]_{SN} = 0  \;, \hspace{40pt}
[X,f]_{SN} = X(f) \;,  \hspace{40pt} 
[X,Y]_{SN} = [X,Y] \;,
}
with $[\ , \ ]$ denoting the Lie bracket.
However, it extends uniquely to arbitrary alternating multi-vectorfields by 
the following relations
\eq{
\label{Rechenregeln}
[X,Y\wedge Z]_{SN} &= [X,Y]_{SN} \wedge Z \;+\; (-1)^{(|X|-1)|Y|}\;Y\wedge[X,Z]_{SN} \;,\\
[X,Y]_{SN} &=-(-1)^{(|X|-1)(|Y|-1)}[Y,X]_{SN} \;,
}
with $|X|$ denoting the degree of $X$.
Thus the degree of the resulting multi-vectorfield is $|[X,Y]_{SN}|=|X|+|Y|-1$. These properties define a so-called \emph{Gerstenhaber-algebra}.
In addition,  the graded Jacobi identity is satisfied
\begin{equation}
\label{grad_jacobi}
\bigl[X,[Y,Z]_{SN}\bigr]_{SN} =\bigl[[X,Y]_{SN},Z\bigr]_{SN}\; +\; (-1)^{(|X|-1)(|Y|-1)}\bigl[Y,[X,Z]_{SN}\bigr]_{SN} \;.
\end{equation}

The algebraic properties of this algebra can now be used to derive Bianchi identities for the fluxes.
For a basis of vector fields $\{e_a\}$, the geometric flux $f^c {}_{ab}$ is given by
\begin{equation}
\label{commutbein}
\bigl[e_a,e_b\bigr]_{SN} = \bigl[e_a,e_b \bigr] = f^c {}_{ab}\;  e_c \;,
\end{equation}
which we assume to be vanishing in the following.
In this case and for vanishing $H$-flux, 
it has been proposed in \cite{Grana:2008yw,Halmagyi:2009te,Aldazabal:2011nj,Andriot:2012wx} 
that the $R$-flux 
can be written in terms of a bi-vectorfield 
$\beta := \frac{1}{2}\beta^{ab} e_a \wedge e_b$ as 
\begin{equation}
	R = \bigl[ \beta, \beta \bigr]_{SN} 
	= \beta^{[\underline{a} m}\; \partial_m\, {\beta}^{\underline{bc}]} \,e_a\wedge e_b\wedge e_c
	=\beta^{[\underline{a} m}\; \mathring\nabla_m\, {\beta}^{\underline{bc}]}
	\,e_a\wedge e_b\wedge e_c \;,
\end{equation}
where the Levi-Civita connection drops out.
Using then \eqref{grad_jacobi}, one finds the trivial relation
\begin{equation}
\label{RBianchi}
0= \bigl[\beta,[\beta,\beta]_{SN}\bigr]_{SN} = \bigl[\beta,{R}\bigr]_{SN} \;,
\end{equation}
and evaluating then the right-hand side of  \eqref{RBianchi}, we obtain\,\footnote{This is consistent with  the $R$-flux Bianchi identity
recently derived  in \cite{Andriot:2012wx} in the double field theory 
approach, if one sets there the derivative with respect to the
``winding'' coordinate $\partial_{\tilde x_i}$ to zero.}
\eq{ 
\label{bianchi_rflux}
        \beta^{[\underline{a} m}\; \partial_m\, {R}^{\underline{bcd}]} =
        \frac{3}{2}\, {R}^{[\underline{ab} m}\, \widetilde Q_m{}^{\underline{cd}]}\, ,
}
where we employed the usual definition of the non-geometric $Q$-flux 
$\widetilde Q_a{}^{bc} = \partial_a \beta^{bc}$. 
Note that \eqref{bianchi_rflux}
 contains the partial derivative $\partial_m$ rather than
the covariant derivative $\mathring\nabla_m$. However, 
the relation can be covariantized as
\eq{ 
\label{bianchi_rfluxb}
        \beta^{[\underline{a} m}\; \mathring\nabla_m\, {R}^{\underline{bcd}]} =
        \frac{3}{2}\, {R}^{[\underline{ab} m}\, Q_m{}^{\underline{cd}]} \;,
}
with $Q_a{}^{bc} = \mathring\nabla_a \beta^{bc}$.
The terms containing the Levi-Civita connection cancel due
to the symmetry of the Christoffel symbols.

As will be discussed elsewhere \cite{BDPRnew}, 
this computation can be generalized. For instance, the identity
\eq{
[\beta,[\beta, e_p]_{SN}]_{SN}- \frac{1}{ 2} [R,e_p]_{SN} =0 
}
leads to a Bianchi identity for the $Q$-flux, that is
\eq{
\label{bianchi_qflux}
\beta^{[\underline{a}m}\, \partial_m \widetilde Q_p{}^{\underline{bc}]} - \frac{1}{3!}
\partial_p R^{[abc]} &= \widetilde Q_p{}^{m[\underline{a}}\,
  \widetilde Q_m{}^{\underline{bc}]} \; .
}
Directly covariantizing this identity, one finds an extra curvature term
so that the Bianchi-identity for the $Q$-flux reads
\eq{
\label{bianchi_qflux2}
  \beta^{[\underline{a}m}\, \mathring\nabla_m Q_p{}^{\underline{bc}]} - \frac{1}{3!}
\mathring\nabla_p R^{[abc]} -2 \beta^{[\underline{a}m}  \beta^{\underline{b}n} 
    \mathring R^{\underline{c}]}{}_{nmp}&= Q_p{}^{m[\underline{a}}\,
  Q_m{}^{\underline{bc}]} \, .
}
The definition of the $Q$-flux leads to another trivial identity
\eq{
       \mathring\nabla_{[\underline{a}}\, 
         Q_{\underline{b}]}{}^{[\underline{cd}]}=-
   \mathring R^{[\underline{c}}{}_{m[\underline{ab}]}\,
   \beta^{\underline{d}]m} \;,
} 
which can be used to exchange the indices $(m\leftrightarrow p)$ in the first
term in \eqref{bianchi_qflux2}.
Therefore, using the Levi-Civita connection, we have the following
six Bianchi-identities written in a geometric basis (i.e. for
vanishing geometric flux):
\eq{
\label{bianchilc}
   \mathring \nabla_{[\underline a}\, H_{\underline{bcd}]}&=0\, ,\\
       \mathring\nabla_{[\underline{a}}\, \mathring
         R^{mn}{}_{\underline{bc}]}&=0\, ,\\
       \mathring R^a{}_{[\underline{bcd}]}&=0\, ,\\
    \mathring\nabla_{[\underline{a}}\, 
         Q_{\underline{b}]}{}^{[\underline{cd}]}+
   \mathring R^{[\underline{c}}{}_{m[\underline{ab}]}\,
   \beta^{\underline{d}]m}&=0 \, ,\\
    \beta^{[\underline{a}m}\, \mathring\nabla_p\, Q_m{}^{\underline{bc}]} - \frac{1}{3!}
\mathring\nabla_p R^{[abc]} &= Q_p{}^{m[\underline{a}}\,
  Q_m{}^{\underline{bc}]} \, ,\\
    \beta^{[\underline{a} m}\; \mathring\nabla_m\, {R}^{\underline{bcd}]} &=
        \frac{3}{2}\, {R}^{[\underline{ab} m}\, Q_m{}^{\underline{cd}]} \;.
}
The claim is that, except for the second line in \eqref{bianchilc},
the remaining five Bianchi identities are precisely in one
to one correspondence with the five Bianchi identities in 
\eqref{wrasebianchi}. The inclusion of geometric flux will
be discussed in \cite{BDPRnew}.
Note that the building blocks of these Bianchi identities are
covariant derivatives of the fluxes and quadratic expressions
of the type appearing in a Jacobi identity.


\subsubsection*{Remark}

Let us remark the following.
In the formalism of double-field theory, 
T-duality is  not only a symmetry for special solutions
of string theory but a symmetry of the underlying equations
of motion.
It is thus tempting to speculate that the two rather strong conditions 
\eqref{jacobi} can be considered as 
universal, that is, if they are satisfied all Bianchi identities
for $H$-, $f$-, $Q$- and $R$-flux are fulfilled. 
If this line of thought is correct, it has the following 
implication for the usual theory formulated with only $H$-flux: 
\begin{itemize}
\item[] The Bianchi identities for $f$-, $Q$-, and $R$-fluxes
should be a consequence of the exact string equations of motion (at all
orders of $\alpha'$) for the metric and the Kalb-Ramond field.
\end{itemize}
This means that there should not exist solutions 
violating any of the Bianchi identities in the other 
T-dual frames. In fact, all exact  solutions we are familiar with, such as
Calabi-Yau manifolds or WZW models,  indeed satisfy
these two stronger conditions.


\section{Conclusion}
\label{sec:con}

In this article, we have considered 
 Einstein-Hilbert,  Gauss-Bonnet 
and higher-order Lovelock actions
within the Palatini formalism, 
where torsion is identified dynamically 
with a three-form flux. For the Einstein-Hilbert action, 
this was straightforward, whereas for the
Gauss-Bonnet case the computation became more
involved. In particular,  for consistency  of the Palatini-variational principle the
three-form had to be covariantly constant and had to satisfy  
a Jacobi identity. These conditions are stronger
than the Bianchi identity for the $H$-flux, but we argued
for a conceivable connection to Bianchi identities for  T-dual fluxes.
More concretely, in double-field theory, our new restrictions can be regarded as the universal
conditions guaranteeing that all Bianchi identities
are satisfied. In this respect, we derived  Bianchi identities
for the non-geometric $Q$- and $R$-flux
including covariant derivative and curvature terms.

We also showed that the  no-go theorem
for relating torsional Gauss-Bonnet gravity to
the second order corrections of the bosonic string 
still applies, even if one imposes the stronger
conditions on the three-form flux. 
However, we presented a form of the next to leading order
bosonic string  action
that contained  a sum of two  Gauss-Bonnet terms, one being
the torsional Gauss-Bonnet gravity action and the
other a pure Kalb-Ramond field dependent contribution.

Clearly, there are a number of open questions. 
For instance, it would be desirable to have a  proof for our conjecture that
the Palatini variational principle carries over to all  higher-order Lovelock actions. 
The relation to string theory also demands the dilaton to be considered. 
Furthermore, more  evidence  for our observation 
that in a T-duality covariant theory one needs to require 
the three-form to be covariantly constant
and to satisfy a Jacobi identity is needed.
Finally, a very interesting question is whether for the case of $R$-flux there is a relation to the 
nonassociativity observed
in  \cite{Bouwknegt:2004ap,Blumenhagen:2010hj,Lust:2010iy,Blumenhagen:2011ph}.


\vskip1cm

\subsubsection*{Acknowledgements}
R.B. would like to thank the KITPC Beijing for hospitality.
E.P. is supported by the Netherlands Organization for Scientific Research (NWO) under the VICI grant 680-47-603.


\clearpage
\nocite{*}
\bibliography{rev}  
\bibliographystyle{utphys}


\end{document}